# Re-Symmetrizing the Broken Symmetry with Isotropic Hydrostatic Pressure

Ke Huang,[1,#] Pengjie Wang,[1] L.N. Pfeiffer,[2] K.W. West,[2] K.W. Baldwin,[2] Yang Liu [1,#,*] and Xi Lin,[1,3,*]

**Affiliation**

[1]International Center for Quantum Materials, Peking University, Haidian, Beijing, China, 100871

[2]Department of Electrical Engineering, Princeton University, Princeton, New Jersey 08544

[3]CAS Center for Excellence in Topological Quantum Computation, University of Chinese Academy of Sciences, Beijing 100190, China

# These authors contribute equally;

* To whom correspondence should be addressed: liuyang02@pku.edu.cn and xilin@pku.edu.cn

February 16, 2019

**Recent progresses in condensed matter physics, such as graphene, topological insulator and Weyl semimetal, often origin from the specific topological symmetries of their lattice structures. Quantum states with different degrees of freedom, e.g. spin, valley, layer, etc., arise from these symmetries, and the coherent superpositions of these states form multiple energy subbands. The pseudospin, a concept analogy to the Dirac spinor matrices, is a successful description of such multi-subband systems. When the electron-electron interaction dominates, many-body quantum phases arise. They usually have digitized pseudospin polarizations and exhibit sharp phase transitions at certain universal critical pseudospin energy splittings. In this manuscript, we present our remarkable discovery of hydrostatic pressure induced degeneracy between the two lowest Landau levels. This degeneracy is evidenced by the pseudospin polarization transitions of the fragile correlated quantum liquid phases near Landau level filling factor $\nu = 3/2$. Benefitted from the**

**constant hole concentration and the sensitive nature of these transitions, we can study the fine-tuning effect of the hydrostatic pressure of the order of 10 μeV, well beyond the meV-level state-of-the-art resolution of other techniques.**

In quantum physics, the conservation laws and energy degeneracies outcome from various symmetries of the system. For example, inversion and rotational symmetries protect the valley degeneracy; and states with opposite orbital angular momenta are degenerate when the time reversal symmetry exists. Energy subbands, especially their fine structures stemming from the pseudospin degree of freedom, can be adjusted by symmetry-breaking effects. In 2D systems, the in-plane magnetic field torques the spin orientation and tunes the Zeeman splitting[1-6]; uniaxial strain breaks the rotational symmetry and changes the valley polarization [7-10]. On the other hand, a moderate hydrostatic pressure of a few kbars conserves the lattice symmetry and usually has negligible effect (μeV-level) on the pseudospin degree of freedom[11, 12].

Pseudospins can stabilize multi-component phases and are of great importance in the study of many-body systems. A strong perpendicular magnetic field $B_\perp$ quantizes the 2D particles' kinetic energy into a set of discrete Landau levels, and gives rise to integer quantum Hall effect when the Landau level filling factor $\nu = nh/eB$ is close to an integer [13]. At very low temperatures, and if disorder is low, the Coulomb interaction leads to the incompressible fractional quantum Hall states which are stable predominantly at odd-denominator fractional $\nu$ [14–17]. These incompressible quantum Hall liquid phases are signaled by vanishing longitudinal resistance $R_{xx}$ and quantized Hall resistance $R_{xy}$. 2D systems with extra pseudospin degrees of freedom have additional sets of Landau levels[6, 7, 18, 19]. When two $N = 0$ Landau levels with different pseudospin flavors approach each other, the multi-component fractional quantum Hall states form and they exhibit pseudospin polarization transitions as one tunes the pseudospin energy splitting $\mathscr{E}$ [1-6, 9, 10, 20-25].

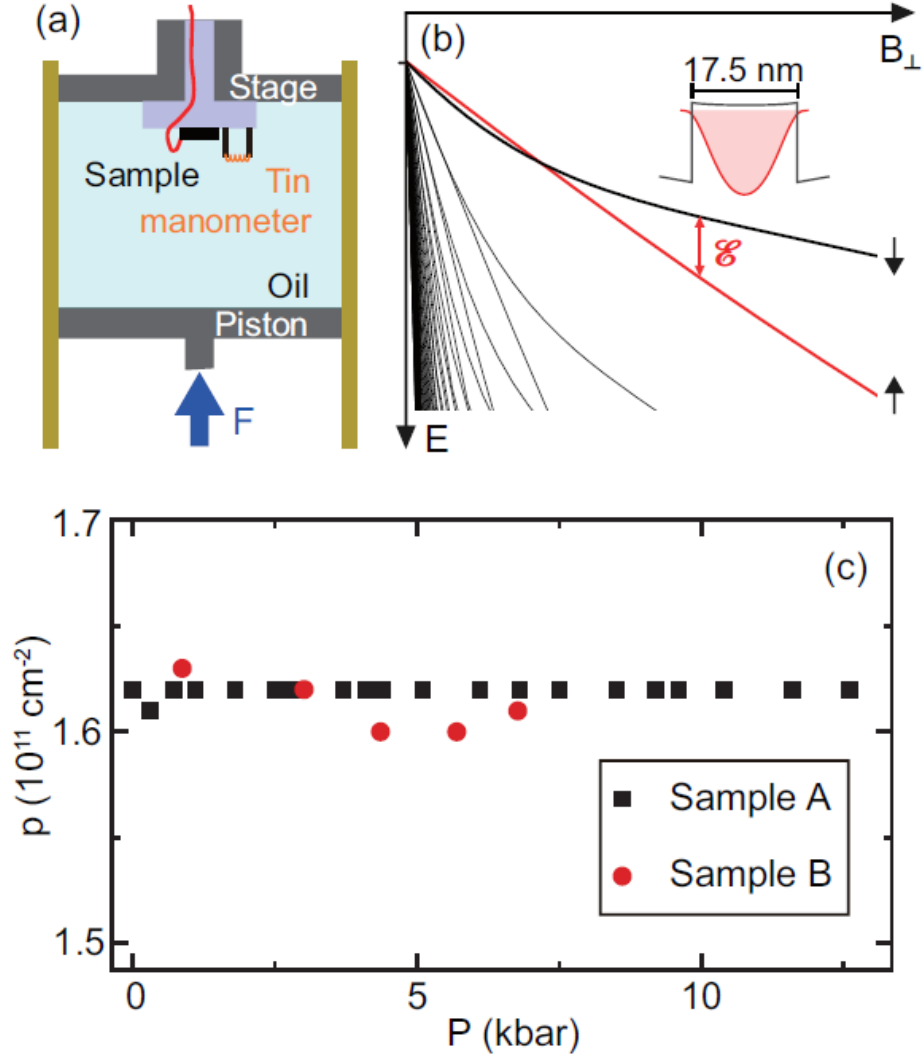

**Fig. 1 Introduction. (a) Experimental setup of the hydrostatic pressure cell. (b) Typical Landau level diagram of 2D holes confined in symmetric, narrow quantum wells. Due to the strong SOI, the Landau levels are non-linear with $B$. The character of the lowest two Landau levels (the thick black and red curves), which are relevant to our observations, are predominately the $N = 0$ harmonic oscillators with ↑ and ↓ pseudospins. (c) The measured**

**2D hole concentration vs. the hydrostatic pressure *P* from both samples reveal almost no density drift in our measurement.**

Here we report experimental discovery of unexpected pseudospin degeneracy in 2D hole systems (2DHSs) confined in symmetric quantum wells. For the first time, we achieve constant density, high quality 2D systems that exhibit pronounced fractional quantum Hall states at large hydrostatic pressure *P* up to about 13 kbar. The isotropic hydrostatic pressure, which only changes the lattice constant but does not vary its geometric symmetry, induces sharp phase transitions between states with digitized pseudospin polarizations. From these transitions, we resolve the pressure-induced 10-μeV-level effect on the pseudospin splitting. These observations indicate that the two lowest Landau levels with opposite spins are nearly degenerate at large hydrostatic pressure *P* > 6 kbar and their energy separation is no larger than a few μeV. This degeneracy suggests that the broken bulk inversion symmetry of the GaAs lattice may be restored by the hydrostatic pressure.

In Figure 2a and b, the magneto-resistances ($R_{xx}$ and $R_{xy}$) and the $\nu$ = 4/3 and 5/3 fractional quantum Hall states' excitation gaps ($^{4/3}\Delta$ and $^{5/3}\Delta$) highlight our discovery. At small hydrostatic pressure *P* = 0.3 kbar, the fractional quantum Hall states at $\nu$ = 4/3 and 5/3 are strong and the 7/5 and 8/5 states are weak, which is a signature of single-component fractional quantum Hall states[17]. When we increase *P*, the $\nu$ = 4/3 state weakens at *P* = 2 kbar and restrengthens at larger *P* > 6 kbar. Meanwhile, the $\nu$ = 5/3 state starts strong at *P* < 2 kbar. It weakens monotonically and saturates at *P* > 6 kbar when $^{5/3}\Delta$ is about 1/3 of its low-*P* value. All these transitions take place between 0 and 6 kbar, and we do not see convincing evidence for further evolution as we increase pressure up to about P = 12.6 kbar. It is worth mentioning that we do not see any signature of transitions at integer fillings ν = 1 and 2 up to the high temperature limit of our system (about 1 K).

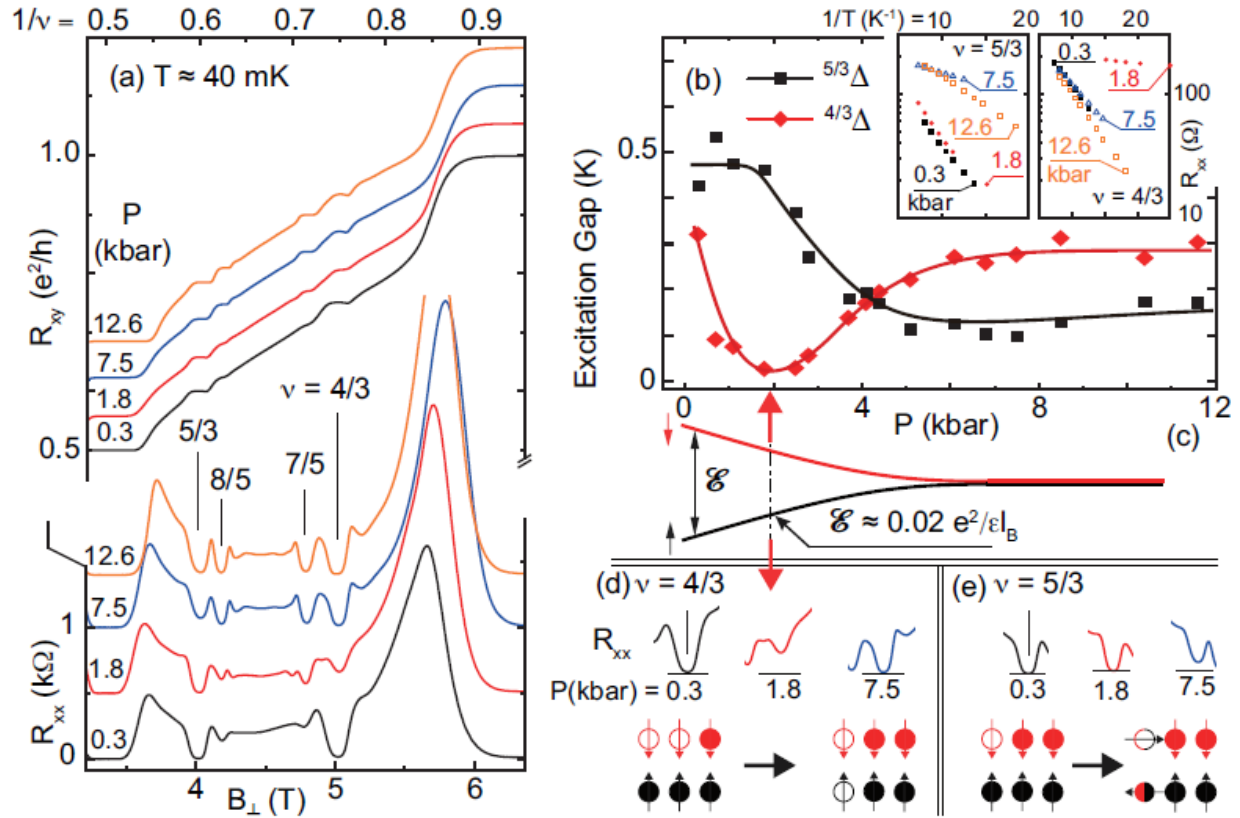

**Fig. 2 Sample A data.** (a) The longitudinal ($R_{xx}$) and Hall resistance ($R_{xy}$) measured from sample A at different hydrostatic pressures *P*. (b) The excitation gap ($^\nu\Delta$) of the ν = 4/3 and 5/3 fractional quantum Hall states (deduced from the $R_{xx}$ vs. *T* data; see inset plots) evolves as a function of *P*, which is consistent with vanishing energy separation $\mathscr{E}$ between the two lowest Landau levels as *P* increases; see panel (c). (d) & (e) $R_{xx}$ minima at ν = 4/3 and 5/3 at different *P* and the cartoon charts to explain our discovery. We mark the $R_{xx}$ = 0 by the thin bar. The ν = 4/3 minimum is strong at *P* = 0.3 and 7.5 kbar, corresponding to pseudospin polarized (left cartoon) and unpolarized (right cartoon) fractional quantum

**Hall states, respectively. The polarization transition appears at $P = 1.8$ kbar, seen as shallower minimum, when $\mathscr{E} = 0.02$ $e^2/4\pi\varepsilon\ell_B$ [21]; $\ell_B$ is the magnetic length. Meanwhile, the $\nu = 5/3$ state could be ferromagnetic [10]. Its minimum continuously weakens as the state transforms from an ↑-pseudospin-polarized state into a coherent superposition of the ↑- and ↓-pseudospin-polarized state when $\mathscr{E}$ vanishes; the left and right cartoons in panel (e), respectively.**

A tentative explanation of the above transitions is that the hydrostatic pressure changes the energy separation $\mathscr{E}$ between the lowest two Landau levels. In periodic lattice without inversion center, e.g. the Zinc blend lattice, the particles' spin orientation relates to its momentum direction through the spin-orbit interaction (SOI), so that the spin degeneracy splits[26]. In bulk GaAs, the SOI splits the spin $S = 3/2$ and $S = 1/2$ hole bands at $k = 0$ (the $\Gamma$-point). The symmetric quantum well confinement along the (001) ($z$-) direction breaks the translational symmetry. The heavy-hole ($|S, S_z\rangle = |3/2, \pm 3/2\rangle$) subbands become lower in energy than the light-hole ($|S, S_z\rangle = |3/2, \pm 1/2\rangle$) subbands due to their larger effective mass along the $z$-direction. When such 2DHS is subjected in a strong perpendicular magnetic field $B_\perp$, the holes' orbital motion is quenched into a set of harmonic oscillators. The SOI mixes hole states with different orbital and spin indices and gives rise to a complex set of Landau levels; see Fig. 1b. The two lowest Landau levels, which are referred to as the two pseudospins, have predominantly $N = 0$ spatial charge distribution but opposite $S_z = \pm 3/2$. The broken inversion asymmetry elevates their energy separation $\mathscr{E}$ through the Dresselhaus effect[26], resulting in fully pseudospin polarized quantum phases at $P = 0$.

The pseudospin splitting 𝓔 vanishes at large hydrostatic pressure at $P > 6$ kbar, evidenced by the factor-of-3 reduction of $^{5/3}\Delta$ from its low-$P$ value. Such weakening is expected to happen only when the pseudospins are nearly degenerate. The Fig. 3b data, measured from another sample at P = 6.6 kbar, confirm this scenario. The fractional quantum Hall effects are weak at odd-nominator fillings ν = 5/3 and 7/5, and are strong at even-nominator fillings ν = 4/3, 8/5 and 10/7. It is unambiguous that the pseudospins are degenerate and 𝓔 is no larger than a few μeV[10, 23, 27]. The ν = 4/3 fractional quantum Hall effect destabilizes and $^{4/3}\Delta$ vanishes at intermediate $P$ = 1.8 kbar. This is also consistent with the expected first-order transition between the pseudospin fully-polarized and unpolarized phases [1, 2, 9, 20, 23]; see Fig. 2c.

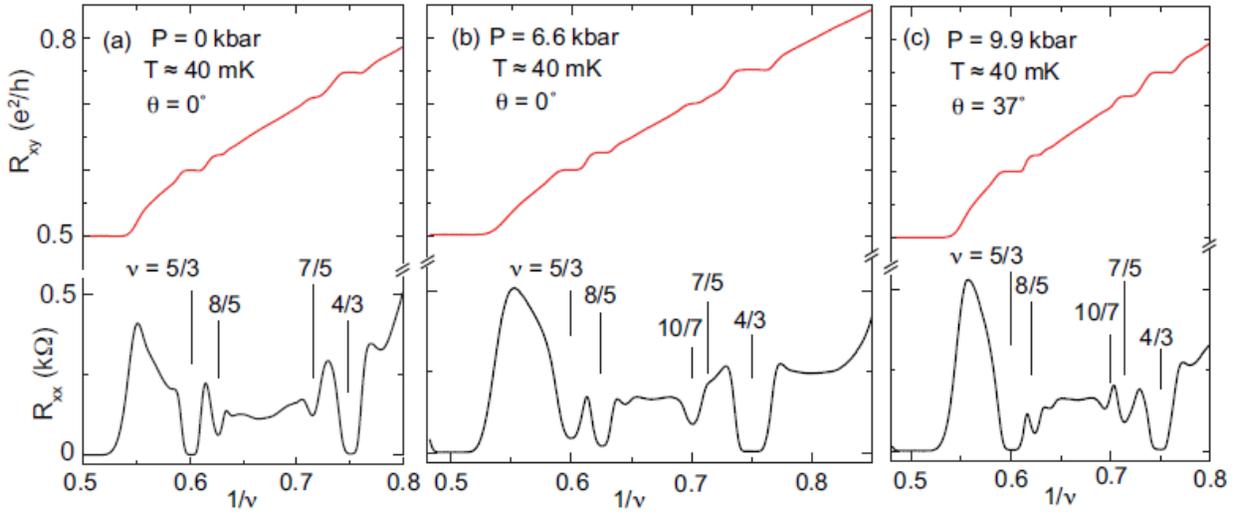

**Fig. 3 Sample B data. (a) & (b) $R_{xx}$ and $R_{xy}$ measured from sample B at θ =0° when $P = 0$ and 6.6 kbar, respectively. (c) Data taken at $P$ =9.9 kbar and θ = 37° resembles features seen in (a), suggesting that the finite θ compensates the $P$-effect.**

The degeneracy of pseudospins at large P is destroyed if we tilt the magnetic field away from the highly-symmetric (001) direction by an angle θ. In Fig. 3c, data taken at $\theta = 37°$ and $P = 9.9$ kbar resemble the features in Fig. 3(a) at $\theta = 0°$ and $P = 0$ when the system is pseudospin fully-polarized: the $R_{xx}$ minima are deep at $v = 4/3$ and $5/3$ and shallow at $v = 7/5$ and $8/5$. Note that the effective Lander g-factor is finite if the magnetic field is not along the (001) and (111) crystal direction. The fact that tilting magnetic field can split the pseudospin degeneracy indicates that the hydrostatic pressure induced degeneracy is rather fragile.

Our study is the first clear evidence that the hydrostatic pressure has a direct impact on the subband structure. The transitions seen in our study can provide a semi-quantitative estimation of its tuning effect. Firstly, Fig. 3c data indicate that $\theta = 37°$ can (over) compensate the effect of 9.9 kbar pressure. Although the accurate calculation is not available, one can estimate that the finite-P and -θ effects are in the same order of 100 μeV. Secondly, the quantum Hall effect at $v = 4/3$ experiences pseudospin polarization transition at $P = 1.8$ kbar when $\mathscr{E}$ is about $0.02 \frac{e^2}{4\pi\epsilon l_B} = 200$ μeV [9, 21], $l_B$ is the magnetic length. Therefore, the hydrostatic pressure controllably tunes the pseudospin splitting at 10 μeV-level. It is only through the sensitive many-body phase transitions that one can probe such fine tuning effect.

It is poorly understood how the hydrostatic pressure can restore the pseudospin degeneracy in 2DHSs. To the leading order, the g-factor remains untouched and the isotropic hydrostatic pressure does not vary the lattice geometric symmetries. Since the finite pseudospin splitting $\mathscr{E}$ is primarily caused by the SOI effect, it is likely that large hydrostatic pressure eliminates the SOI effect in 2DHSs and restores the degeneracy. This effect is extremely challenging to either explore or estimate with other approaches: The angle-resolved photoemission spectroscopy (ARPES) has a state-of-the-art resolution of sub-meV-level [28, 29](28, 29); First principle theoretical calculations based on many-body perturbation theory, on the other hand, resolve band gaps of at most in-the-order-of 0.1 eV [30-32]. Recent studies in 2D electron systems have seen broken rotational symmetry at large hydrostatic pressure. Unfortunately, the pressure-effect is indecisive because of the significant carrier density reduction[11, 12, 33]. Secondly, although both samples show weakening of the quantum Hall effect at $\nu = 5/3$, there are differences between data from the two neighboring samples. While the Fig.3b data is perfectly consistent with collapsing $\mathscr{E}$, the $\nu = 7/5$ fractional quantum Hall state which is expected to be weak appears strong at high pressure in Fig. 2a. It is possible that the residue structural asymmetry of the quantum well confinement along the z-direction strengthens this state.

## Method

The two samples used in this report are made from the same GaAs wafer grown by molecular beam epitaxy along the (001) direction. The wafer consists of a 17.5-nm-wide GaAs quantum well bounded on either side symmetrically by undoped $Al_{0.3}Ga_{0.7}As$ spacer layers and carbon δ-doped layers. The as-grown density of these samples is $p = 1.6 \times 10^{11}$ cm$^{-2}$ and the low-temperature ($T = 0.3$ K) mobility is above 100 m$^2$/Vs. Each sample has a van der Pauw geometry, with alloyed InZn contacts at the four corners of a $2 \times 2$ mm$^2$ piece cleaved from the wafer. We mount the sample on the epoxy sample stage and fill the pressure cell with oil. We press the piston at room temperature to apply the hydrostatic pressure, and then measure low-temperature $P$ via the superconducting critical temperature (about 3 K) of the in-situ Tin manometer. The commercial high-pressure cell is installed in a sample probe of Leiden CF-CS81-600, and can be cooled downed to less than 40 mK. The base temperature of the dilution refrigerator is less than 8 mK and the base temperature of the sample probe is below 25 mK. We use low-frequency (< 50 Hz) lock-in technique to measure the transport coefficients.

## Acknowledgments

We acknowledge support by the National Key Research and Development Program of China (Grant No. 2015CB921101 and No. 2017YFA0303301), NSFC (Grant No. 11674009) and the Key Research Program of the Strategic Priority Research Program of Chinese Academy of Sciences (Grant No. XDB28000000) for measurements, and the NSF (Grants DMR-1305691, and MRSEC DMR-1420541), the Gordon and Betty Moore Foundation (Grant GBMF4420), and Keck Foundation for the material growth and characterization. We thank R. Winkler for providing the charge distribution, potentials and the calculated Landau level fan diagram in Fig. 1(b). We thank M. Shayegan, Fuchun Zhang, Hao Zhang, Ji Feng, Xinzheng Li, Xiongjun Liu and Yan Zhang for valuable discussion.

## Author contributions

K.H. and P.W. performed the measurements. K.H., P.W. and X. L. prepared the low temperature high pressure cell setup. L.N.P., K.W.W. and K.W.B prepared and supplied the GaAs wafer. Y.L. and X.L. initiated the research. K.H., P.W., Y.L. and X.L. analyzed the results. K.H., Y.L. and X.L. discussed the results and wrote the manuscript.

## Additional information

## Competing financial interests

The authors declare no competing financial interests.